\documentclass[prb,aps,twocolumn,amsmath,amssymb,floatfix,
superscriptaddress]{revtex4}

\usepackage[dvips]{graphics}
\usepackage{color}
\definecolor{dred}{rgb}{0.75,0,0}

\begin{document}

\title{High degree of current rectification at nanoscale level}

\author{Madhumita Saha}

\affiliation{Physics and Applied Mathematics Unit, Indian Statistical
Institute, 203 Barrackpore Trunk Road, Kolkata-700 108, India}

\author{Santanu K. Maiti}

\email{santanu.maiti@isical.ac.in}

\affiliation{Physics and Applied Mathematics Unit, Indian Statistical
Institute, 203 Barrackpore Trunk Road, Kolkata-700 108, India}

\begin{abstract}

We address an unexpectedly large rectification using a simple quantum 
wire with correlated site potentials. The external electric field, 
associated with voltage bias, leads to unequal charge currents for two 
different polarities of external bias and this effect is further enhanced 
by incorporating the asymmetry in wire-to-electrode coupling. Our 
calculations suggest that in some cases almost cent percent rectification 
is obtained for a wide bias window. This performance is valid against 
disorder configurations and thus we can expect an experimental verification 
of our theoretical analysis in near future.

\end{abstract}

\maketitle

\section{Introduction}

Designing of an efficient rectifier at nanoscale level has been the subject
of intense research after the prediction of molecular rectifier by Aviram
and Ratner~\cite{r1} in $1974$. Following this pioneering work, interest in 
this area has rapidly picked up with several theoretical propositions and 
experimental verifications and most of these works involve small organic 
molecules with donor-acceptor pair between metallic electrodes~\cite{r2,r3,
r4,r5,r55,r6,r7,r8,r9,r10,r11,r12}. Recently a DNA-based rectifier has 
also been established~\cite{r13} which exhibits a large rectification 
ratio of about $15$ at $1.1\,$V.

To achieve rectification (viz, $I(V)\ne I(-V)$), energy levels of the 
bridging material have to be aligned differently for the positive and 
\begin{figure}[ht]
{\centering \resizebox*{7.5cm}{2cm}{\includegraphics{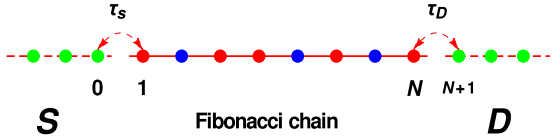}}\par}
\caption{(Color online). Quasi-periodic (Fibonacci, $5$th generation i.e.,
$8$ atomic sites) chain attached to $1$D source (S) and drain (D) 
electrodes. The filled red and blue circles correspond to two different 
lattices (say) $A$ and $B$, respectively.}
\label{f1}
\end{figure} 
negative biases. This can be done in two different ways: (i) placing an
asymmetric conductor between source and drain electrodes~\cite{r9,r11}, 
keeping identical conductor-to-electrode couplings, and (ii) considering a 
symmetric conductor with unequal conductor-electrode couplings~\cite{r12}. 
The understanding is that for both these two cases resonant energy levels, 
in presence of finite bias, are arranged distinctly for two different biased 
conditions which results finite rectification. Therefore, when both these 
two conditions are satisfied one can expect maximum rectification. 

So far mostly molecular systems~\cite{r6,r7,r8,r9,r10,r11,r12} have been 
used to design rectifying diodes, but constructing them using single molecules
is a major challenge~\cite{janes} and still many open questions remain which 
certainly demand further study. Therefore, {\em designing of a rectifier 
using simple geometric structure which provides high rectification ratio 
is a matter of great interest.} In the present work we 
essentially focus on it and make an attempt to establish that a simple 
$1$D chain with correlated site potentials can exhibit a very high degree 
of rectification, and sometimes it becomes nearly close to $\pm 100\%$ for 
a wide bias window. 
This performance is valid against disordered configurations which we confirm 
by comparing 
the results of different $1$D quasi-periodic chains like Fibonacci (Fibo), 
Thou-Morse (TM), Copper-mean (CM) and Bronze-mean (BM) and all these 
systems are constructed by using two primary lattices, namely $A$ and $B$, 
following the specific inflation rules~\cite{quasiall,tm,skmfib}. For the 
case of
Fibonacci chain the rule is: $A \rightarrow AB$ and $B \rightarrow A$. 
Therefore, applying successively this substitutional rule, staring with 
$A$ or $B$ lattice we can construct the full lattice chain for any 
particular generation, say $p$-th generation, obeying the prescription 
$F_p=F_{p-1} \otimes F_{p-2}$. So, if we start with $A$ lattice then the
first few generations of the Fibonacci series are $A$, $AB$, $ABA$, 
$ABAAB$, $ABAABABA$, $\dots$, etc. The inflation rules for the other three
quasiperiodic chains which we consider here i.e., TM, CM and BM are: 
$A \rightarrow AB$, $B \rightarrow BA$; $A \rightarrow ABB$, 
$B \rightarrow A$, and $A \rightarrow AAAB$, $B \rightarrow A$, respectively. 
Using these rules we construct the quasiperiodic chains for any desired 
generation starting with any lattice site $A$ or $B$.

The rest of the paper is arranged as follows. In Sec. II we present the
model and theoretical framework for calculations. The results are described
in Sec. III, and finally, in Sec. IV we conclude our findings.

\section{Model and theoretical framework}

The calculations are worked out using wave-guide theory based on 
tight-binding (TB) framework. In this framework the Hamiltonian of the full
system, schematically shown in Fig.~\ref{f1}, can be written as
$H=H_{\mbox{\tiny el}} + H_{\mbox{\tiny ch}} + H_{\mbox{\tiny tn}}$, where
$H_{\mbox{\tiny el}}$, $H_{\mbox{\tiny ch}}$ and $H_{\mbox{\tiny tn}}$
correspond to the Hamiltonians of the electrodes (source and drain), 
quasi-periodic chain and chain-to-electrode tunneling coupling, 
respectively. In terms of on-site potential $\epsilon_i$ and nearest-neighbor
hopping integral $t$, the TB Hamiltonian of the chain can be written as:
\begin{equation}
H_{\mbox{\tiny ch}}=\sum\limits_{i=1} \epsilon_i c_i^{\dagger} c_i + 
\sum\limits_{i=1} t \left(c_{i+1}^{\dagger} c_i + c_i^{\dagger} c_{i+1} 
\right)
\label{equ1}
\end{equation}
where $c_i^{\dagger}$ ($c_i$) represents the electronic creation 
\begin{figure}[ht]
{\centering \resizebox*{6.5cm}{4cm}{\includegraphics{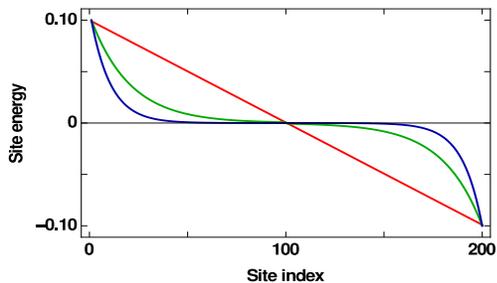}}\par}
\caption{(Color online). Voltage dependent site energies of a $200$-site 
chain for three different electrostatic potential profiles, shown by three
distinct colored curves, at the bias voltage $V=0.2\,$V.} 
\label{f2}
\end{figure}
(annihilation) operator. In a similar way we can write the TB Hamiltonian
$H_{\mbox{\tiny el}}$ of the two side-attached perfect $1$D electrodes
parameterized by $\epsilon_0$ and $t_0$. These electrodes are coupled to
the sites $1$ and $N$ of the conductor (described by $H_{\mbox{\tiny tn}}$)
through the coupling parameters $\tau_S$ and $\tau_D$ (see Fig.~\ref{f1}),
where $N$ being the total number of lattice sites of the bridging conductor.

In presence of a finite bias $V$ between the source and drain, an electric 
field is developed across the chain and hence its site energies are voltage
dependent~\cite{el1,el2}. It gives $\epsilon_i=\epsilon_i^0 + \epsilon_i(V)$, 
where $\epsilon_i^0$ is voltage independent and it becomes $\epsilon_A$ or 
$\epsilon_B$ depending on the lattice sites $A$ or $B$. The dependence
of $\epsilon_i(V)$ is associated with electron screening as well as
bare electric field at the junction. In the absence of any screening 
electric field is uniform across the junction~\cite{el1,el2} which makes 
$\epsilon_i(V)=V/2-iV/(N+1)$ (linear variation, red line of Fig.~\ref{f2}) 
for a $N$-site chain.
Whereas, long-range electron screening makes the profile non-linear
as shown by the green and blue curves of Fig.~\ref{f2}. In our calculations
we consider these three different potential profiles to have a complete
idea about the bare and the screened electric field profiles, and their 
effects on rectification as in realistic case different materials possess
different electron screening which will yield different field variations.
For a slight variation from these potential profiles no significant change
is observed in the physical properties, and thus our findings may be 
implemented in realistic cases.

To evaluate transmission probability across the conducting junction we solve
a set of coupled linear equations containing wave amplitudes of distinct
lattice sites of the chain~\cite{wg,skm}. Assuming a plane wave incidence, 
we can write
the wave amplitude at any site $n$ of the source as $A_n=e^{ikn}+re^{-ikn}$,
where $k$ is the wave-vector and $r$ being the reflection coefficient. 
While, for the drain electrode we get $B_n=\tau e^{ikn}$, where $\tau$
represents the transmission coefficient. For each $k$, associated with
the injecting electron energy, we find the transmission probability from
the expression $T(E)=|B_1|^2=|\tau|^2$. Once it is determined, the net 
junction current for a particular bias voltage $V$ at absolute zero 
temperature is obtained from the relation~\cite{datta}
\begin{equation}
I(V) = \frac{e}{\pi \hbar} \int\limits_{E_F-\frac{eV}{2}}^{E_F+\frac{eV}{2}}
T(E) \, dE
\label{eqcurr}
\end{equation}
where $E_F$ represents the Fermi energy. Finally, we define the rectification 
ratio as~\cite{r13} $RR=|I(V)|/|I(-V)|$. $RR=1$ suggests no rectification.
In all calculations we set the common parameter values as: $t=1\,$eV, 
$\epsilon_0=0$, $t_0=3\,$eV and, unless otherwise stated $E_F=0$.

\section{Results and discussion}

\begin{figure}[ht]
{\centering \resizebox*{7.5cm}{6.5cm}{\includegraphics{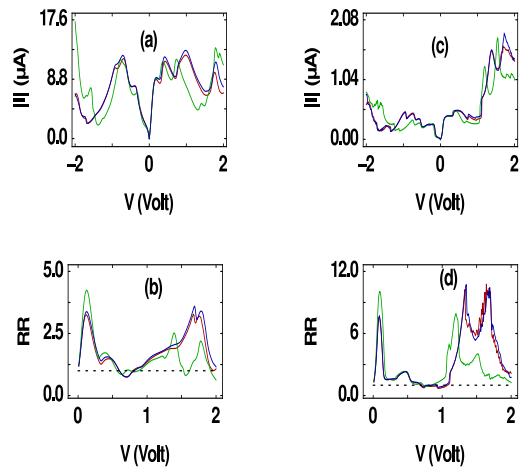}}\par}
\caption{(Color online). $|I|$ and $RR$ as a function of voltage for a 
$8$th generation ($N=34$) Fibonacci chain with 
$-\epsilon_A=\epsilon_B=0.5\,$eV considering both linear (red curve) and 
two non-linear (green and blue curves) potential profiles. In the first 
column the results are shown for the symmetric coupling 
($\tau_S=\tau_D=1\,$eV), while in the second column they are presented 
for the asymmetric coupling ($\tau_S=0.1\,$eV and $\tau_D=1\,$eV). Dashed
line corresponds to $RR=1$.}
\label{f3}
\end{figure}

Now we present our results. In Fig.~\ref{f3} we show the variations of 
$|I|$ both for the forward and reverse biased conditions along with the
rectification ratio $RR$ considering three different electrostatic potential
profiles where the first and second columns correspond to the symmetric and
asymmetric wire-to-electrode couplings, respectively. Two observations are
noteworthy. First, the currents for the positive and negative biases are
quite close to each other in the limit of symmetric coupling 
(Fig.~\ref{f3}(a)), whereas {\em they differ distinctively in the case of 
asymmetric coupling} (Fig.~\ref{f3}(c)) though the currents in this case are 
much less than the previous one. Second, for a wide bias window 
($\sim 1$-$2\,$V) the rectification ratio is significantly large, in the 
limit of asymmetric coupling, reaching a maximum of $\sim 11$. This is 
a reasonably large value compared to the reported results for different 
conducting junctions considering both symmetric molecular structures as 
well as asymmetric molecule-to-lead couplings where RR varies between 
$2$ to $10$~\cite{r9,r12}.

\begin{figure}[ht]
{\centering \resizebox*{7.75cm}{6.5cm}{\includegraphics{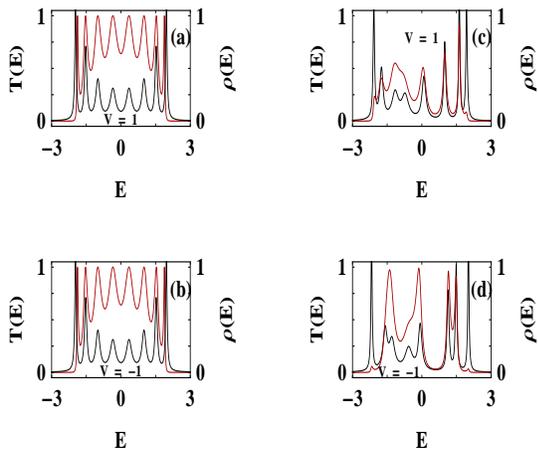}}\par}
\caption{(Color online). Transmission probability $T(E)$ (red color) as 
a function of energy $E$ for ordered ($\epsilon_i=0 \, \forall \, i$) and 
Fibonacci chains ($-\epsilon_A=\epsilon_B=0.5\,$eV) considering a linear 
bias drop. The average density of states $\rho(E)$ (black color) is 
superimposed in each spectrum. Here we set $\tau_S=\tau_D=1\,$eV and $N=8$.}
\label{ados}
\end{figure}

To illustrate the mechanism of rectification let us focus on the spectra
given in Fig.~\ref{ados}. For the fully perfect chain ($\epsilon_i^0=0\,
\forall \,i$), $T$-$E$ spectrum in presence of positive bias (red curve of 
Fig.~\ref{ados}(a)) exactly matches with what we get in the case of negative 
bias (red curve of Fig.~\ref{ados}(b)). Therefore, for this junction 
identical currents are obtained upon the integration of the transmission
function $T$, resulting a vanishing rectification. While, comparing the
spectra given in Figs.~\ref{ados}(c) and (d), worked out for the positive 
and negative biases considering a Fibonacci chain, it is clearly seen that 
the transmission spectra differ sharply which results a finite rectification.
To achieve rectification the essential thing is that, as stated earlier, the 
energy levels of the bridging material have to be aligned differently for
the positive and negative biases. The alignments of energy levels for the 
two different wires are clearly reflected from the $\rho$-$E$ spectra
(black curves of Fig.~\ref{ados}). Thus increasing the misalignment higher 
$RR$ is expected and it can be done further by including additional asymmetric 
factors like asymmetric environmental effects~\cite{env}, inconsistent 
gating~\cite{gat}, etc.

\begin{figure}[ht]
{\centering \resizebox*{7cm}{9cm}{\includegraphics{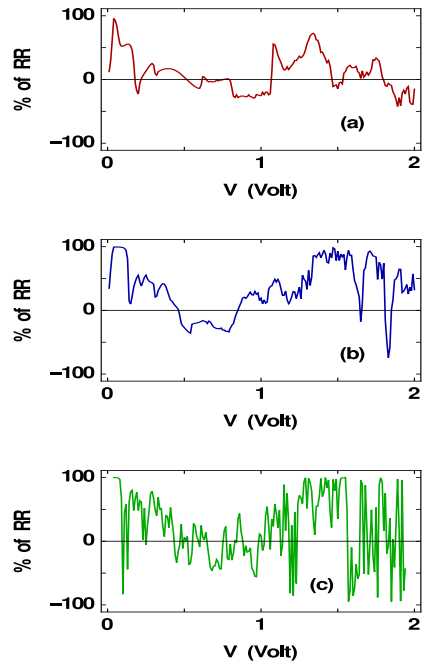}}\par}
\caption{(Color online). Percentage of $RR$ as a function of bias voltage 
$V$ for three different sizes of the Fibonacci chain considering a linear
bias drop, where (a), (b) and (c) correspond to $N=55$, $89$, $233$, 
respectively. Here we choose $\tau_S=0.1\,$eV and $\tau_D=1\,$eV, and 
$-\epsilon_A=\epsilon_B=0.5\,$eV.}
\label{ratio}
\end{figure}

Figure~\ref{ratio} shows the percentage of rectification (defined as
$(|I(V)|-|I(-V)|)/(|I(V)|+|I(-V)|)=(RR-1)/(RR+1)$) as a function of 
voltage for three 
different sizes of the Fibonacci chain. Quite interestingly we see that 
for {\em wide voltage regions nearly cent percent rectification is 
obtained}, and thus the present system can be utilized as a perfect 
rectifier.

\begin{figure}[ht]
{\centering \resizebox*{6cm}{4cm}{\includegraphics{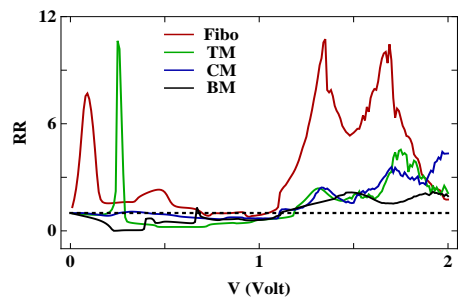}}\par}
\caption{(Color online). $RR$-$V$ characteristics for four different 
quasi-periodic chains with inconsistent wire-to-electrode coupling
($\tau_S=0.1\,$eV and $\tau_D=1\,$eV) in presence of linear potential 
profile. Here we set $N=34$, $32$, $43$ and $43$ for Fibo, TM, CU and BM 
chains, respectively, and choose $-\epsilon_A=\epsilon_B=0.5\,$eV.}
\label{f4}
\end{figure}

In order to justify the robustness of rectification against disorderness,
in Fig.~\ref{f4} we present $RR$-$V$ characteristics for some typical 
quasi-periodic lattices. All these junctions provide finite rectification  
where $RR$ varies in a wide range, and from these curves it can be 
emphasized that any one of such lattices can be used to achieve 
the goal of rectification action. Looking carefully into the spectrum
(Fig.~\ref{f4}) it is observed that $RR$ reaches to zero ($\sim -100\%$) 
for a reasonable voltage window ($\sim 0.23$-$0.38\,$V) for the junction 
containing BM wire. One could also get opposite scenario i.e., 
$\sim + 100\%$ rectification through any one of these junctions. This 
is solely associated with the interplay between the arrangements of 
lattice sites and electrostatic potential profile. 

\begin{figure}[ht]
{\centering \resizebox*{6.5cm}{6cm}{\includegraphics{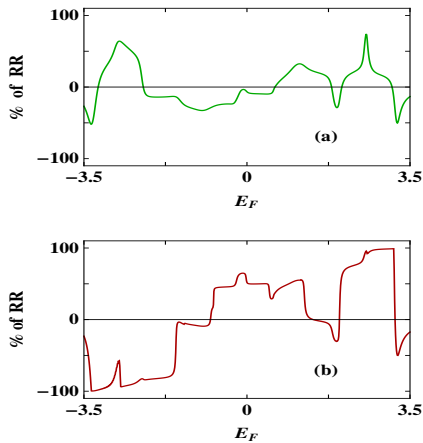}}\par}
\caption{(Color online). Percentage of $RR$ as a function of Fermi energy
$E_F$ for a $5$th generation Fibonacci chain 
($-\epsilon_A=\epsilon_B=0.5\,$eV) considering a linear bias drop at $V=2\,$V,
where (a) and (b) correspond to the symmetric ($\tau_S=\tau_D=1\,$eV) and
asymmetric ($\tau_S=0.1\,$eV and $\tau_D=1\,$eV) wire-to-electrode couplings,
respectively.}
\label{f5}
\end{figure}

Finally, in Fig.~\ref{f5} we discuss the possibilities of regulating the 
rectification ratio {\em externally} for a fixed bias voltage. This can be 
achieved by tuning the Fermi energy, which on the other hand is controlled 
by external gate voltage. It is worthy to note that upon changing $E_F$ one 
can get a wide variation of $RR$ (viz, $-100\%$ to $+100\%$), and thus it 
can be emphasized that the present model can be utilized to get 
{\em externally controlled} rectifier at nano-scale level.

Before the end, we would like to point out that, apart from rectifying 
action all these quantum wires characterized by quasi-periodic lattices 
show another uncommon property of junction current where an increase in 
the bias voltage results a reduction of net current (see Figs.~\ref{f3} 
and \ref{f4}). This is the so-called negative differential conductance 
(NDC) effect~\cite{ndc}, and its detailed analysis will be given in our 
forthcoming paper.

\section{Concluding remarks}

To conclude, in the present work we have attempted to establish a model
quantum system that can exhibit a high degree of current rectification 
at nanoscale level. An unexpectedly large rectification ratio has been 
obtained, and most importantly, we see that in some cases nearly cent 
percent rectification can be achieved for a wide bias window. Our results 
are also valid against disordered configurations which we have confirmed 
by considering different kinds of disordered systems. Though the proposition
given here is based on purely theoretical arguments, we hope that its 
experimental verification can be done in near future. 

Finally, we want to note that although the results have been worked out at 
zero temperature, all the physical features remain invariant at finite 
temperature ($\sim 300$ K) as thermal broadening of energy levels is much 
weaker than the broadening caused as a result of wire-to-electrode 
couplings~\cite{datta}.

\section{Acknowledgment}

MS would like to acknowledge University Grants Commission (UGC) of India
for her research fellowship.

\end{document}